\documentclass[namedreferences]{SolarPhysics}
\usepackage[optionalrh,solaenum]{spr-sola-addons} 
\usepackage{graphicx}        
\usepackage{color}           
\usepackage{url}             

\newcommand{\be}{\begin{equation}}
\newcommand{\ee}{\end{equation}}
\newcommand{\bex}{\begin{equation}\notag}
\newcommand{\eex}{\end{equation}\notag}
\newcommand{\bea}{\begin{eqnarray}}
\newcommand{\eea}{\end{eqnarray}}
\newcommand{\beax}{\begin{eqnarray*}}
\newcommand{\eeax}{\end{eqnarray*}}
\newcommand{\ba}{\begin{array}}
\newcommand{\ea}{\end{array}}

\newcommand{\grad}{ {\bf \nabla } }

\newcommand{\vecB}{{\mathbf B}}
\newcommand{\vecE}{{\mathbf E}}
\newcommand{\vecU}{{\mathbf U}}

\newcommand{\vecS}{{\mathbf S}}

\newcommand{\vecv}{{\mathbf v}}

\newcommand{\vecxhat}{{\mathbf {\hat x}}}
\newcommand{\vecyhat}{{\mathbf {\hat y}}}
\newcommand{\veczhat}{{\mathbf {\hat z}}}
\newcommand{\scrJ}{{\mathcal{J}}}
\newcommand{\scrB}{{\mathcal{B}}}

\begin{document}

\begin{article}

\begin{opening}

\title{Can We Determine Electric Fields and Poynting Fluxes
from Vector Magnetograms and Doppler Measurements?}

\author{G.H.~\surname{Fisher}$^{1}$\sep
        B.T.~\surname{Welsch}$^{1}$\sep
        W.P.~\surname{Abbett}$^{1}$      
       }
\runningauthor{G.H. Fisher \textit{et al.}}
\runningtitle{Electric Fields and Poynting Fluxes}

   \institute{$^{1}$ Space Sciences Laborary, UC Berkeley, CA, USA\\
                     email: \url{fisher@ssl.berkeley.edu}, \url{welsch@ssl.berkeley.edu}, \url{abbett@ssl.berkeley.edu}
             }

\begin{abstract}
The availability of vector-magnetogram sequences with sufficient accuracy and
cadence 
to estimate the temporal derivative of the magnetic field allows 
us to use Faraday's law to find an approximate solution for the 
electric field in the photosphere, using a Poloidal--Toroidal 
Decomposition (PTD) of the magnetic field and its partial time derivative.  
Without additional information, however, the electric field found from 
this technique is under-determined -- Faraday's law provides no information 
about the electric field that can be derived the gradient of a scalar 
potential.  Here, we show how additional information in the form of 
line-of-sight Doppler-flow measurements, and motions transverse to 
the line-of-sight determined with \textit{ad-hoc} 
methods such as local correlation 
tracking, can be combined with the PTD solutions to provide much more 
accurate solutions for the solar electric field, and therefore the 
Poynting flux of electromagnetic energy in the solar photosphere.
Reliable, accurate maps of the Poynting flux are essential for quantitative
studies of the buildup of magnetic energy before flares and coronal mass
ejections.
\end{abstract}
\end{opening}

\section{Introduction}
\label{Introduction} 

The launch of SDO, with its ability to measure the Sun's vector 
magnetic field anywhere on the disk with a high temporal cadence,
promises to usher in a new era of solar astronomy.  This new era of
measurement demands new approaches for the analysis and use of this data.  
We show in this
article how the vector magnetic field and Doppler-flow measurements that
can now be made with HMI (\opencite{Scherrer2005})
lead to new methods for determining
the electric field vector, and the Poynting Flux vector 
\be
\vecS = {1 \over 4 \pi} c \vecE \times \vecB
\label{equation:sdef}
\ee
at 
the solar photosphere. The Poynting flux measures the flow of 
electromagnetic energy
at the layers where the magnetic field is determined.  
Quantitative observational studies of how energy flows into the corona 
depend on deriving accurate estimates of the Poynting flux.

Most work estimating the Sun's electric field or Poynting flux
either explicitly or implicitly assumes that the electric field is determined by
ideal MHD processes, and therefore the problem can be reduced to determining 
a velocity field associated with the observed magnetic-field evolution.  
One class of velocity estimation techniques are ``Local Correlation 
Tracking'' (LCT) methods, which essentially capture pattern 
motions of the line-of-sight magnetic
field or white-light intensity.  This approach was pioneered by 
\inlinecite{November1988}.  Other implementations include
the Lockheed--Martin LCT code (\opencite{Title1995};
\opencite{Hurlburt1995}), 
``Balltracking'' (\opencite{Potts2004}), and the FLCT code 
(\opencite{Fisher2008}).  Another class of velocity-estimation methods 
incorporate solutions of the vertical component of the magnetic
induction equation into determinations of the velocity field
(\opencite{Kusano2002}; \opencite{Welsch2004}; \opencite{Longcope2004}; 
Schuck, 2006, 2008;
\opencite{Chae2008}).
The work we present in this article incorporates 
solutions of the three-dimensional
magnetic induction equation, using the electric field as 
the fundamental variable, rather than the velocity field.

The temporal evolution of the Sun's magnetic field is governed by
Faraday's law,
\be
{ \partial \vecB \over \partial t} = - \grad \times c \vecE\ .
\label{equation:faraday}
\ee
If one can can make a map on the photosphere of 
$\partial \vecB / \partial t$, can one determine $\vecE$ by uncurling this
equation?  Addressing this question was the focus of \inlinecite{Fisher2010a},
in which a poloidal--toroidal decomposition (PTD) of the temporal derivative
of the magnetic field was used to invert Faraday's law to find $\vecE$.
\inlinecite{Fisher2010a} found that 
one could indeed find solutions for $\vecE$ that
solve all three components of Faraday's law, but the solutions are not unique:
the gradient of a scalar function can be added to the PTD solutions for 
$\vecE$ without affecting $\grad \times \vecE$.  \inlinecite{Fisher2010a} 
explored
two different methods for determining the scalar function 
using \textit{ad-hoc} and variational methods, both of which enforced the assumption,
from ideal MHD, that $\vecE$ must be normal to $\vecB$.  Unfortunately, the
agreement with a 
test case from an MHD solution, while better than conventional 
correlation-tracking methods, was still disappointing.
The authors concluded that including additional information
from other observed data was one possible approach for improving the
electric field inversions.

In this article, we use the same MHD simulation test case used in 
\inlinecite{Welsch2007} and \inlinecite{Fisher2010a} to 
show that using Doppler-flow 
measurements to determine the electric scalar potential, especially in regions
where the magnetic field is primarily horizontal, can dramatically improve the
inversion for the electric field and the Poynting flux.

In Section \ref{section:ptd} we review the PTD formalism that describes how one
can derive the purely inductive part of the electric field from measurements
that estimate the time derivative of $\vecB$, and the technique of Section 3.2
of \inlinecite{Fisher2010a}, showing how one can derive a potential electric
field, which, when added to the inductive part of the electric field,
is normal to the magnetic field.  This is useful in
generating electric-field solutions that are both consistent with
Faraday's law and 
with ideal MHD, which is generally believed to be a good approximation 
in the solar photosphere.

Section \ref{section:doppler} argues from physical grounds why magnetic-flux
emergence may make a large contribution to the part of the electric field
attributable to a potential function.  Then, starting from this argument,
we derive a Poisson equation for an electric-field potential function that
is determined primarily from knowledge of the vertical velocity field, as
determined from Doppler measurements, and
the horizontal magnetic field near polarity inversion lines where the field
is nearly horizontal.  The electric field from this contribution
is then added to that determined from the PTD solutions.  
We then apply this technique to the MHD simulation test
data, to compare the electric field from the simulation with that from PTD 
alone, and with that from combining PTD with Doppler measurements.

In Section \ref{section:flct}, we try a similar approach, but instead of using
contributions to the horizontal electric field from Doppler measurements, we
use non-inductive contributions to the electric field
determined from the FLCT correlation-tracking 
technique, applicable in regions where the magnetic field is mainly
vertical.  This technique is essentially the three-dimensional 
analogue of the ILCT technique described by \inlinecite{Welsch2004}.  We also
try combining PTD with contributions from both 
the Doppler measurements and those from
FLCT, and compare with the simulation data.

Our results are summarized in Section \ref{section:conclusions}, along with
a discussion of where additional work is needed.

\section{Poloidal--Toroidal Decomposition}
\label{section:ptd}

Here, we present only a brief synopsis of the PTD method of deriving an electric
field $\vecE$ that obeys Faraday's law.  
More detail can be found in Section 2 of \inlinecite{Fisher2010a}.

Since the three-dimensional magnetic field vector is a solenoidal quantity, 
one can express the magnetic field in terms of two scalar functions, $\scrB$
(the ``poloidal'' potential)
and $\scrJ$ (the ``toroidal'' potential), as follows:
\be
\vecB = \grad \times \grad \times \scrB \veczhat + \grad \times \scrJ \veczhat 
\ .
\label{equation:ptdstatic}
\ee
Taking the partial time derivative of Equation (\ref{equation:ptdstatic})
one finds
\be
\dot \vecB = \grad \times \grad \times \dot \scrB \veczhat 
+ \grad \times \dot \scrJ \veczhat \ .
\label{equation:ptde}
\ee
Here, the overdot denotes a partial time derivative.  We will now assume
a locally Cartesian coordinate system, in which the directions parallel to
the photosphere are denoted with a ``horizontal'' subscript $h$, 
and the vertical direction is denoted with subscript $z$.  One can then
re-write Equations (\ref{equation:ptdstatic}) and (\ref{equation:ptde}) 
in terms of horizontal and vertical derivatives as
\be
\vecB = \grad_h \left( \partial \scrB \over \partial z \right) 
+ \grad_h \times \scrJ \veczhat
- \nabla_h^2 \scrB \veczhat , 
\label{equation:ptdstatichz}
\ee
and
\be
\dot \vecB = \grad_h \left( \partial \dot \scrB \over \partial z \right) 
+ \grad_h \times \dot \scrJ \veczhat
- \nabla_h^2 \dot \scrB \veczhat . 
\label{equation:ptdehz}
\ee

One useful property of the poloidal--toroidal decomposition is that the
scalar functions $\dot \scrB$, $\dot \scrJ$, and 
$\partial \dot \scrB / \partial z$ 
can all be determined by knowing the time derivative of the magnetic-field
vector in the plane of the photosphere.  By examining the $z$-component of
Equation (\ref{equation:ptdehz}), the $z$-component of the curl of
Equation (\ref{equation:ptdehz}), and the horizontal divergence of Equation
(\ref{equation:ptdehz}), one can derive the following three two-dimensional
Poisson equations for $\dot \scrB$, $\dot \scrJ$, and
$\partial \dot \scrB / \partial z$:
\be
\nabla_h^2 \dot \scrB = -\dot B_z\ ,
\label{equation:poissonbz}
\ee
\be
\nabla_h^2 \dot \scrJ = -( 4 \pi / c ) \dot J_z = - \veczhat 
\cdot ( \grad \times \dot \vecB_h ) ,
\label{equation:poissonjz}
\ee
and
\be
\nabla_h^2 ( {\partial \dot \scrB / \partial z} ) = \grad_h \cdot \dot \vecB_h .
\label{equation:poissondivbh}
\ee
Here, $\dot B_z$ and $\dot \vecB_h$ denote the partial time derivatives of the
vertical and horizontal components of the magnetic field, respectively.
Solving these three Poisson equations provides sufficient information
to determine an electric field that satisfies Faraday's law.

By comparing the form of Equation (\ref{equation:faraday})
with Equations (\ref{equation:ptde}) and (\ref{equation:ptdehz})
it is clear the following must be true:
\bea
\grad \times c \vecE 
&=& - \grad \times \grad \times \dot \scrB \veczhat - \grad \times \dot \scrJ 
\veczhat
\label{equation:curleptd} \\
&=& - \grad_h (\partial \dot \scrB / \partial z ) - 
\grad_h \times \dot \scrJ \veczhat + \nabla_h^2 \dot \scrB \veczhat.
\label{equation:curleptdexpand}
\eea
Uncurling Equation (\ref{equation:curleptd}) yields this expression for the
electric field $\vecE$:
\be
c \vecE = - \grad \times \dot \scrB \veczhat - \dot \scrJ \veczhat - 
c \grad \psi \equiv c \vecE^I - c \grad \psi .
\label{equation:eptd}
\ee
Here, $-\grad \psi$ is the contribution to the electric field from a scalar
potential, for which solutions to Faraday's law reveal no information.
The solution for $\vecE$ without the contribution from $-\grad \psi$,
$\vecE^I$, is the purely inductive solution determined from the PTD method.  
Within this article, this solution will be referred to simply as the PTD
solution or the PTD electric field.  Note that 
the PTD solution is not unique.
While solutions for
$\partial \dot \scrB / \partial z$ are necessary to ensure that Faraday's
law is obeyed, the PTD solution for the electric field itself depends
only on $\dot \scrB$ and $\dot \scrJ$.
This means that the PTD electric field
is the same for distributions of $\dot B_z$ and $\dot \vecB_h$ 
which have differing
values of $\grad_h \cdot \dot \vecB_h$, but the same values of 
$( \grad_h \times \dot \vecB_h ) \cdot \veczhat$ and $\dot B_z$.
Thus the PTD solutions for $\vecE^I$ are under-determined. 

\inlinecite{Fisher2010a} described two techniques for deriving an 
electric-field contribution from a scalar potential, 
in an effort to resolve the
under-determined nature of the PTD solutions.  The first technique, described
in Section 3.2 of that article,
presents an \textit{ad-hoc} iterative method for deriving a
scalar potential electric field which, when added to the PTD solution, 
results in an electric field that is normal to $\vecB$, and hence consistent
with ideal MHD.  The second technique, based on a variational method, 
finds a scalar potential electric field that, when added to the PTD solution,
minimizes the area integral of $|\vecE|^2$ or $|\vecv|^2$.  When compared
to the original electric field from the simulation test case, the iterative
method applied to the PTD solutions 
showed a qualitative consistency, but not detailed agreement
with the simulation electric field, 
while the electric field computed with the variational technique showed 
poor agreement.  \inlinecite{Fisher2010a} concluded that significant
improvement in the agreement of the inverted electric field with the 
real electric field requires additional
observational information beyond the temporal evolution of $\vecB$.

\section{The Importance of Doppler Flow Measurements to the Electric Field}
\label{section:doppler}

\begin{figure}    
   \centerline{\includegraphics[width=1.0\textwidth,clip=]{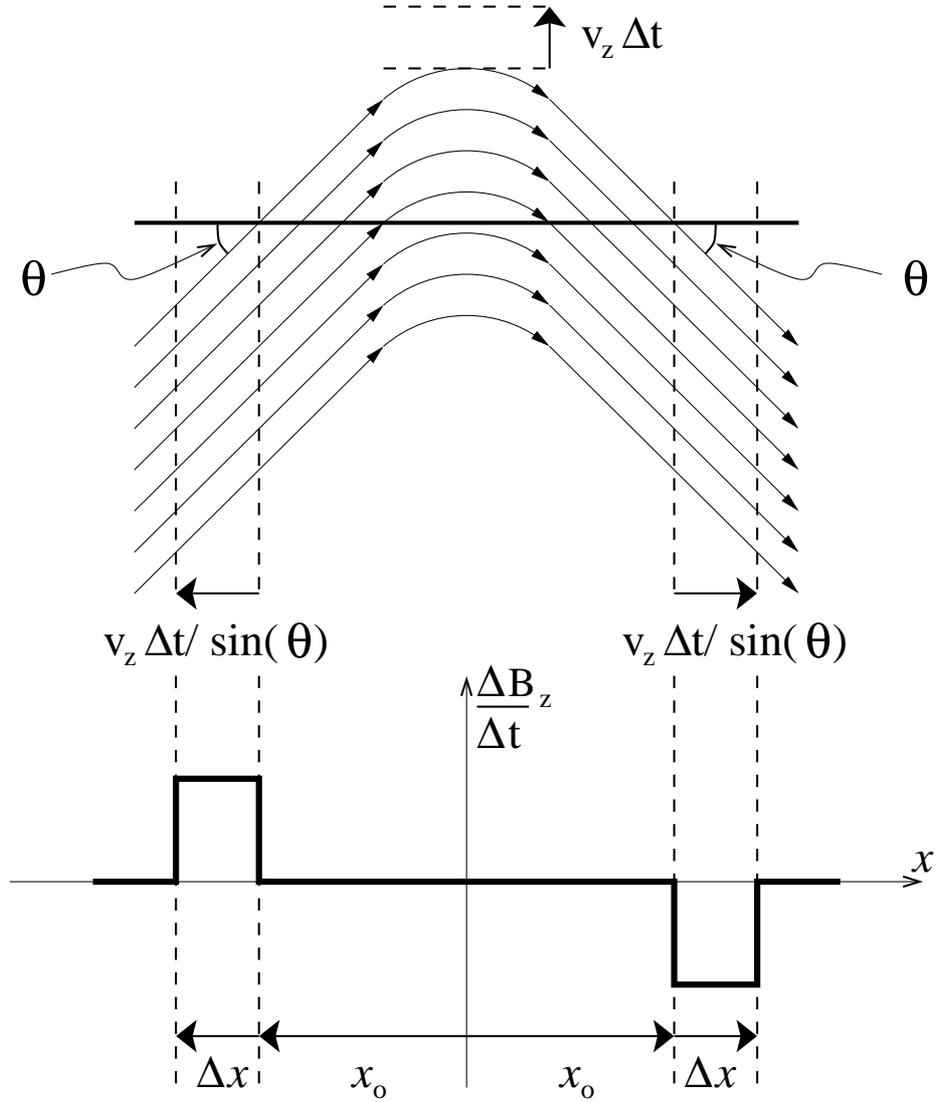}
              }
   \caption{
   Schematic illustration of the emergence of new flux over 
   a time interval
   $\Delta t$, viewed in a vertical plane normal to the polarity inversion
   line (PIL) in an idealized bipolar flux system. The emerging flux is
   rising at a speed $v_z$, which could be inferred by the Doppler shift
   measured by an observer viewing the PIL from above. The width of the
   bipolar flux system (the distance from the outer edge of one pole to the
   outer edge of the other pole) at the beginning of $\Delta t$ is $2x_0$.
   Notice that the change in $B_z$ at the outer edges of the emerging flux
   region is large, while the change in $B_z$ at the PIL itself --- where the
   flux is actually emerging --- is zero (see text).
   }
   \label{figure:pilemerge}
   \end{figure}

We argue here that when flux emergence occurs, 
much of the missing information about non-inductive
contributions to the electric field
is contained in Doppler flow information (see also \opencite{Ravindra2008}
and \opencite{Schuck2010}), 
particularly near polarity
inversion lines (PILs), where 
the horizontal magnetic field is much stronger than
the vertical field.
We illustrate this point with a simple thought-experiment, shown
schematically in Figure \ref{figure:pilemerge}.
Consider the emergence of new magnetic flux in an idealized bipolar-flux
system, where the PIL maintains its orientation as flux continuously
emerges from below the photosphere. Imagine that vector magnetogram and
Doppler observations are taken from a vantage point normal to the solar
surface.  Let us focus attention on what is happening near the center of
the PIL. Suppose the magnetic field there remains time-invariant as flux
continues to emerge, so the time derivative of the magnetic field there is
zero, implying that Faraday's law cannot be used to infer the physics of the
emerging flux. Yet the electric field at this location should be very
large, driven by the upward motion of the plasma carrying the strong,
horizontal field. In this case, magnetic-flux emergence will have a strong
inductive signature at the edges of the idealized active region, where the
vertical magnetic field is changing rapidly, but not near the center of
the PIL. Thus, it seems plausible that the electric field near PILs in
more realistic emerging-flux configurations will have a significant
non-inductive component.

Starting from this perspective, we have explored enhancements to the 
PTD method that use Doppler flow information to more tightly constrain 
the PTD electric field solutions, with the additional assumption that the
photospheric electric field is primarily governed by ideal MHD processes.  
Directly above PILs, the vertical velocity and the observed horizontal 
component of the magnetic field unambiguously determine the horizontal 
electric field:
\be
c \vecE^D_h = -v_z \veczhat \times \vecB_h\ ,  
\label{equation:pilelec}
\ee
where we assume that $|B_z| / |B_h|$ is small.
If we 
can use line-of-sight Doppler velocity measurements to estimate $v_z$, 
we add a powerful constraint to the PTD solution for the electric field.  
Of course, we would like to use the Doppler information away from PILs 
as well, but are hindered by two complications: {\it i}) flows parallel to 
the magnetic field will not affect the electric field at all, but may 
contribute to the observed Doppler velocity signal, and {\it ii}) when the 
vertical component of the magnetic field $[B_z]$ becomes significant 
compared to the horizontal field $[B_h]$, there is an additional contribution 
to the horizontal electric field from flow parallel to the surface, which 
is not accounted for.

We now develop a formal solution for a non-inductive
contribution to the electric field that includes information from 
Doppler-shift measurements, and apply it to 
a test case with a known electric field.
First, from the pair of synthetic vector magnetograms taken from the
ANMHD simulation test case described in \inlinecite{Welsch2007} and
Section 3.1 of \inlinecite{Fisher2010a}, we use the PTD method
to find an electric-field solution, neglecting
any contribution from a scalar electric-field potential function.  
We use the numerical techniques and 
boundary conditions described in Section 3.1 of
\inlinecite{Fisher2010a}.
Second, we compute a candidate horizontal electric field from vertical 
velocities taken from the simulation as synthetic Doppler-flow 
measurements, and horizontal magnetic fields from the synthetic vector 
magnetograms, from Equation (\ref{equation:pilelec}) above.  
This electric field is then multiplied 
by a ``confidence function'', which is near unity at PILs, but decreases 
to zero when $|B_z / B_h |$ is no longer small. This reflects our lack of 
confidence in the accuracy of this horizontal Doppler electric field in those 
locations, for the reasons described earlier.  
The specific form for the confidence function is probably not important.  Here,
we assume the confidence function $w$ is given by
\be
w = {\rm exp} [-(|B_z| / |B_h|)^2 / \sigma^2 ]\ ,
\label{equation:xi}
\ee
where $\sigma$ is a free parameter that can be adjusted, and in the
specific cases shown in this article was set somewhat arbitrarily to $0.6$.
We define the ``modulated'' electric 
field within the plane of the magnetogram as
\be
\vecE^M_h = w \vecE^D_h
\label{equation:emodulated}
\ee
Third, we take the divergence of this modulated horizontal 
electric field $\vecE^M_h$, and find the electric-potential function that 
can best represent it by setting 
\be
c \vecE^{\chi} = - \grad_h \chi\ , 
\label{equation:chidef}
\ee
where 
$\chi$ solves the Poisson equation 
\be
\nabla_h^2 \chi = - \grad_h \cdot c \vecE^M_h\ .  
\label{equation:poissondop}
\ee
Because the synthetic vector magnetograms and Doppler flows
taken from the MHD simulation use periodic
boundary conditions, we use FFT techniques to solve
Equation (\ref{equation:poissondop}).
Adding this contribution onto the PTD solutions means 
that information about the electric field at PILs has been incorporated, 
while also maintaining consistency with Faraday's Law, since 
$\vecE^{\chi}$ has no curl.  Since we generally expect ideal MHD to be
a good approximation for conditions in the solar photosphere, we then
remove the components of $\vecE$ parallel to $\vecB$ by adding the electric 
field from a second potential function $\psi$, using the iterative
technique described in Section 3.2 of \inlinecite{Fisher2010a}:
\be
c \vecE^{\mathrm {tot}} = c \vecE^I + c \vecE^{\chi} - \grad \psi\ ,
\label{equation:evectotal}
\ee
where $\grad \psi \cdot \vecB = (c \vecE^I + c \vecE^{\chi} ) \cdot \vecB$.

\begin{figure}    
   \centerline{\includegraphics[width=1.0\textwidth,clip=]{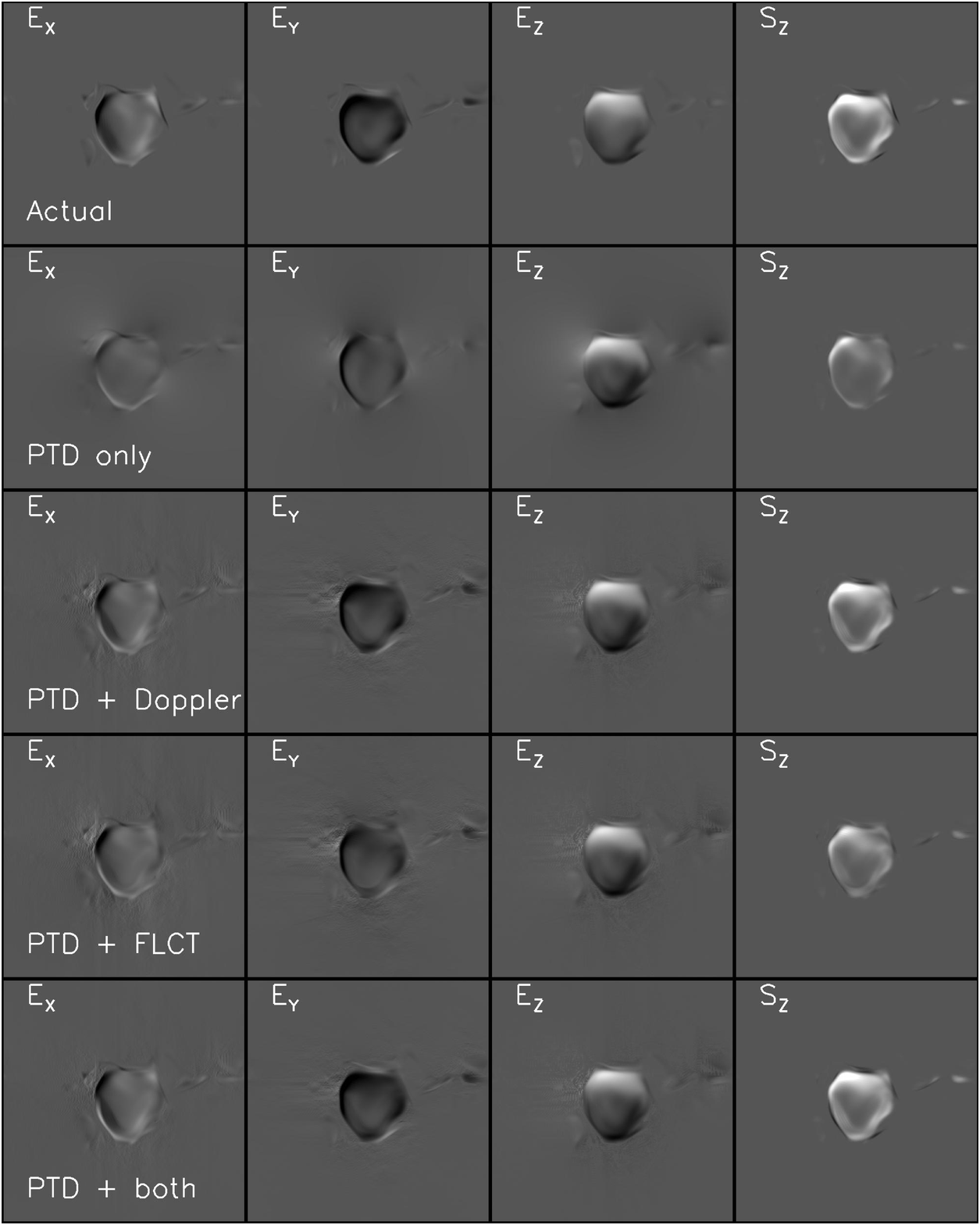}
              }
   \caption{Top row: The 
    three components of the electric field and 
    the vertical Poynting flux from the MHD reference simulation of 
    emerging magnetic flux in a turbulent convection zone.
    Second row: The inductive 
    components of {\bf E} and $S_z$ determined using the PTD 
    method.  Third row: {\bf E} and $S_z$ derived by 
    incorporating Doppler flows around PILs into
    the PTD solutions.  
    Note the dramatic improvement in the estimate
    of $S_z$.  Fourth row: {\bf E} and $S_z$ derived by incorporating 
    only non-inductive FLCT derived flows into the PTD solutions.  Note
    the poorer recovery of $E_x$, $E_y$, and $S_z$ relative to the case
    that included only Doppler flows. Fifth row: {\bf E} and $S_z$ derived
    by including both Doppler flows and non-inductive FLCT flows into the PTD
    solutions.  Note the good recovery of $E_x$, $E_y$, and $S_z$, and the
    reduction in artifacts in the low-field regions for $E_y$ (best viewed
    in the electronic version of the article).
    }
   \label{figure:efieldcomp}
   \end{figure}

\begin{figure}
   \centerline{\includegraphics[width=1.0\textwidth,clip=]{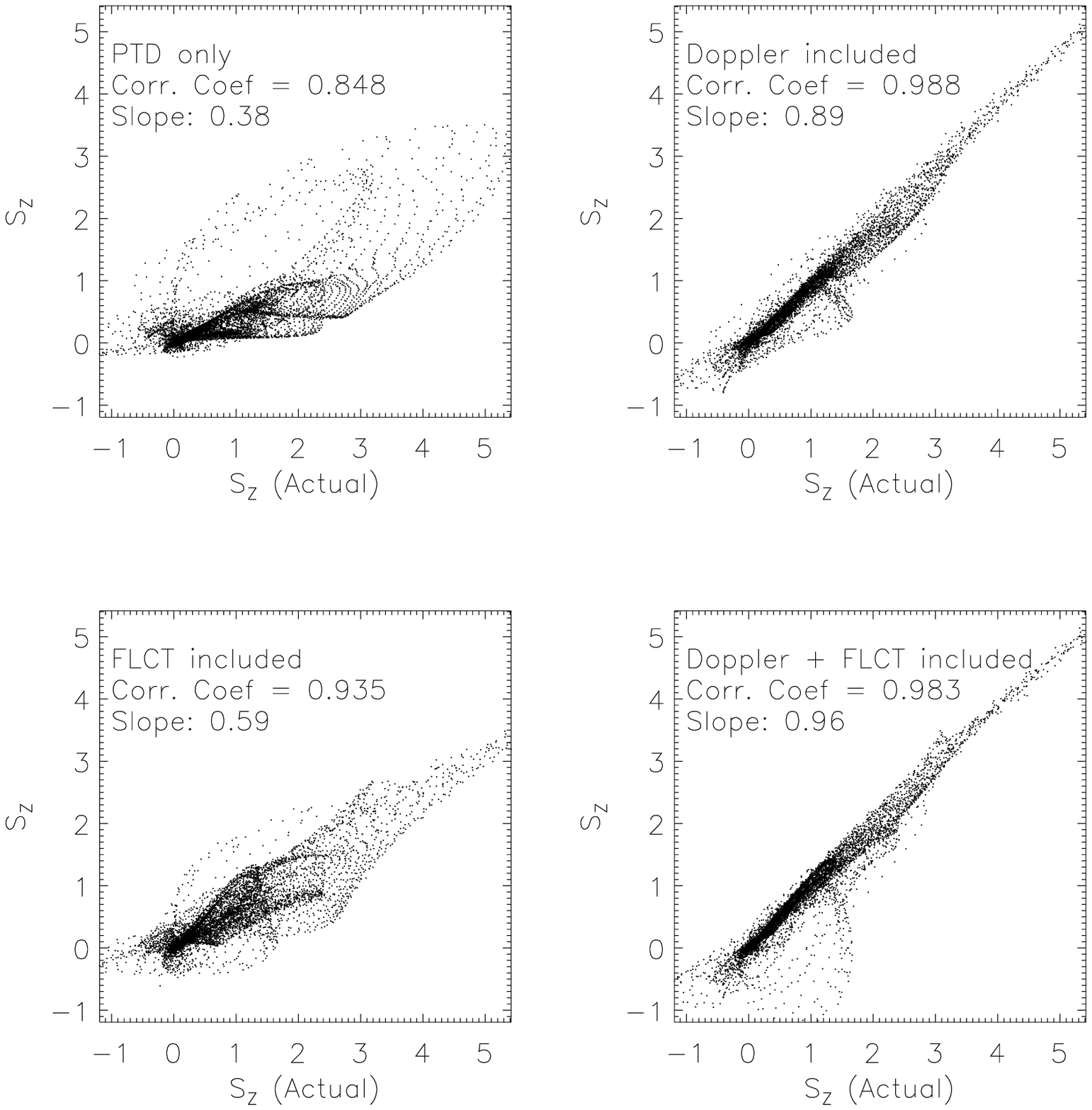}
              }
  \caption{
    Upper left: A comparison of the vertical component of the 
    Poynting flux derived from the PTD method alone with the actual
    Poynting flux of the MHD reference simulation.  Upper right:
    A comparison between the simulated results and the improved
    technique that incorporates information about the vertical flow 
    field around PILs into the PTD solutions. Lower left: Comparison of
    the vertical Poynting flux when non-inductive FLCT-derived flows are
    incorporated into the PTD solutions. Lower right: Comparison of
    the vertical Poynting flux when both Doppler flow information and
    non-inductive FLCT-derived flows are incorporated into the PTD solutions.
    Each scatterplot also shows the computed linear correlation coefficient,
    as well as the slope of the fit derived with IDL's LADFIT function.
    Poynting flux units are in [$10^5\ {\rm G}^2\ {\rm km\ }{\rm s}^{-1}$]
    }
   \label{figure:szscatterplot}
\end{figure}

The resulting solutions for $\vecE$ are shown in the third row of Figure 
\ref{figure:efieldcomp}, with a scatterplot comparison of $S_z$ of the PTD
method and the PTD plus Doppler information with the actual simulation
electric fields shown in 
the top two panels of Figure \ref{figure:szscatterplot}.
These portions of the figures show that the recovery of the electric-field
components and the Poynting flux is dramatically better than PTD alone.

\section{How Important are Horizontal, Non-Inductive Flows?}
\label{section:flct}

In the previous section, we considered the role of Doppler-flow measurements
in determining non-inductive contributions to the horizontal electric field,
and found that combining this information with the PTD solutions for Faraday's
law results in a dramatic improvement in the recovery of the electric field.
However, this treatment neglects possible contributions to the horizontal
electric field away from PILs where a cross product of horizontal velocity
with vertical magnetic field could
also contribute to the horizontal electric field.
Contributions to the horizontal electric field that solve the induction equation
have already been incorporated by the PTD solutions, but as with vertical
velocities, there could be a sub-space of horizontal flows that do not 
contribute to Faraday's law.

To evaluate this effect, we estimate horizontal velocities
using the FLCT local-correlation tracking (LCT) code 
(see \opencite{Fisher2008}), 
available from 
\hbox{\url{http://solarmuri.ssl.berkeley.edu/~fisher/public/software/FLCT/C_VERSIONS/}}
using images of the vertical component of the magnetic field.
Velocities were not computed for pixels with a vertical magnetic field 
strength below
370G (see discussion in \opencite{Welsch2007}), with the windowing parameter
$\sigma$ set to five pixels.  The low-pass filtering option was not invoked.
The result is a map of the apparent horizontal-velocity field
$[ \vecU_h \equiv U_x \vecxhat + U_y \vecyhat ]$
computed at the strong vertical magnetic-field locations, and with
velocities at all other locations set to zero.  A candidate horizontal
electric field is estimated by setting 
\be
c \vecE^{\mathrm {LCT}}_h = -\vecU_h \times \veczhat B_z\ .
\label{equation:vecelctdef}
\ee
To consider only non-inductive contributions from $\vecU_h$, we perform the
same general operation as in the previous section, namely to multiply 
$\vecE^{\mathrm {LCT}}_h$ by
a confidence function, and then eliminate the inductive part of the electric
field.  Here, the confidence function will be the complement of the
confidence function used for the Doppler case, since the LCT estimates
are nearly useless near PILs, where the Doppler results should be reliable,
while the LCT results should be best when the magnetic field is 
mostly vertical (and where the Doppler measurements are useless).

We define $c \vecE^{\zeta} = - \grad_h \zeta$, and assume that
\be
\nabla_h^2 \zeta = - \grad_h \cdot (1-w) c \vecE^{\mathrm {LCT}}_h
\label{equation:poissonlct}
\ee

Once this equation has been solved and 
$\vecE^{\zeta}$ has been computed, it can be added to the PTD solutions
for $\vecE^I$, and as in the previous section, a second potential solution can
be found that eliminates components of $\vecE$ parallel to $\vecB$.  
Note that combining the PTD solutions with $\vecE^{\zeta}$ in this way
is like the approach used in the ILCT technique
described by \inlinecite{Welsch2004}, except that
solutions of a single component of the induction equation are replaced by
solutions to all three components of the induction equation.

The resulting electric field and Poynting flux can be compared
to the actual case, the un-altered PTD case, and the case where only the
Doppler information is used.  
The electric field and Poynting flux results
are shown as the fourth row of panels in Figure \ref{figure:efieldcomp} and
a scatterplot of the Poynting flux values with the actual values is shown
in the lower left panel of Figure \ref{figure:szscatterplot}.
While the overall performance of the FLCT case is better than that of PTD
alone, it is not significantly better than simply applying the iterative
method directly to the PTD results as was described in Section 3.2 of
\inlinecite{Fisher2010a}.  It is definitely not as good as the performance
we show from the Doppler-only case.  We conclude that most of the useful
information about the non-inductive electric field, at least for this
particular simulation of strong flux emergence, is contained within the
Doppler flow information.

Does the LCT information, when added to the Doppler-flow information, 
significantly improve the resulting estimate for the electric field?
To answer this question, we have added both the LCT and Doppler 
electric-field 
information to the PTD solutions, and again found a potential
function to eliminate components of $\vecE$ parallel to $\vecB$.  The resulting
electric field and Poynting-flux maps are shown in the fifth row of panels
in Figure \ref{figure:efieldcomp}, and a scatterplot of the vertical Poynting
flux is shown in the lower-right panel of Figure \ref{figure:szscatterplot}.

The linear 
correlation coefficient in the Poynting-flux scatterplot is not signficantly
improved by adding the LCT results to the Doppler results, but the slope of
the fit (determined by using IDL's {\sf {LADFIT}} function) is somewhat better.
Further, examining the
maps of $E_x$ and $E_y$ show a reduction in artifacts in the behavior of
the recovered electric-field components, compared to the Doppler and LCT
cases.  We conclude that at least for this simulation, which exhibited
strong flux emergence, most of the additional
useful information beyond solutions to Faraday's law
is contained within the Doppler velocity measurements, with some
additional improvement when non-inductive LCT-derived electric fields are added.

Finally, we wish to add a comment about solutions to the PTD equations
themselves.  The PTD solutions used in this article did not use FFT solutions
for $\dot \scrB$ and $\dot \scrJ$, even though the simulations are periodic,
but instead used Neumann boundary conditions for $\dot \scrB$ for the reasons
described in Section 2.2 of \inlinecite{Fisher2010a}.  For the current study, we
compared the results of using FFT solutions of the PTD equations with those
shown in the figures in this article, and found noticeable
degradations in the fits of the model Poynting fluxes to the actual model
values.  If one is interested in the most accurate reconstruction of the
vertical Poynting flux, we recommend not using FFT solutions of the PTD
equations.

\section{Discussion and Conclusions}
\label{section:conclusions}

We have reviewed how the PTD solutions of Faraday's law for $\vecE$ can be
found using temporal sequences of vector magnetograms that can be obtained
with the HMI instrument on NASA's SDO mission.  We discussed why these
solutions are under-determined, and the importance of determining the
contributions to the electric field that can be derived from a scalar potential.

We demonstrate, using simulation data where the true electric
field is known, that knowledge of the vertical-velocity field 
(obtainable by Doppler measurements) can provide important information
about the electric field.  When this information is combined with the
PTD solutions of Faraday's law, dramatically more accurate recovery of the
true electric field is possible.  We find that additional information
about flows from local correlation-tracking methods can also be combined with
the PTD solutions, but the additional information is signficantly less important
than that from the Doppler measurements.  We are able to quantitatively
reconstruct the electromagnetic Poynting flux in the simulations by using
our combination of the PTD solutions and those from Doppler measurements.

This ``proof-of-concept'' demonstration argues strongly for the development
of electric-field and Poynting-flux tools to 
be used routinely in the analysis of HMI vector magnetic-field
measurements.  Routinely available Poynting-flux maps will be useful for
scientific studies of flare-energy buildup, understanding the flow of magnetic
energy in the solar atmosphere prior to CME initiation, and will aid in
understanding the flow of energy that heats the corona.  Further, the PTD
formalism for the magnetic field itself (Equation (\ref{equation:ptdstatichz}))
allows for a straight-forward
decomposition of the Poynting flux into changes
in the potential-field energy, and the flux of free magnetic energy (see
\opencite{Welsch2006} and the end of
Section 2.1 of \opencite{Fisher2010a}).  The flux of free magnetic energy
is especially important in determining how energy builds up in flare-productive
active regions.

To find solutions for $\vecE$ and the Poynting flux $\vecS$ using the PTD
formalism plus Doppler measurements requires only the solution of three
two-dimensional Poisson equations.  While real vector-magnetogram patches
will not have periodic boundary conditions (as were employed in this article),
straightforward numerical techniques exist to solve these equations
routinely.  Preliminary investigations also indicate that generalizing the
PTD solutions and Doppler measurements to cases of non-normal viewing angle
will be straightforward.  In our opinion, the major obstacle that remains
before such solutions can be routinely applied to the HMI data, is
a detailed understanding of how measurement errors and disambiguation errors
in the vector magnetograms will affect the solutions, and how the effects
of these errors are best ameliorated.

\begin{acks}
This research was funded by the NASA Heliophysics Theory Program
(grant NNX08AI56G), the
NASA Living-With-a-Star TR\&T Program (grant NNX08AQ30G), 
by the NSF SHINE program (grants ATM0551084 and ATM0752597), and 
support from NSF's AGS Program (grant ATM0641303) for our participation
in the University of Michigan's CCHM Project.  The authors are grateful
to US taxpayers for providing the funds necessary to perform this work.
The authors wish to acknowledge Dick Canfield for his pioneering work in the
use of vector magnetograms in solar physics.  The inspiration for the
work described here can
be traced to a Solar MURI workshop held at UC Berkeley in 2002, in which
Dick Canfield played a major role in defining long-term research goals for
the use of vector magnetograms in quantitative models of the Sun's atmosphere.
\end{acks}

   
\bibliographystyle{spr-mp-sola}

\bibliography{abbrevs,short_abbrevs,full_lib,bib_mods}  

\IfFileExists{\jobname.bbl}{} {\typeout{}
\typeout{****************************************************}
\typeout{****************************************************}
\typeout{** Please run "bibtex \jobname" to obtain} \typeout{**
the bibliography and then re-run LaTeX} \typeout{** twice to fix
the references !}
\typeout{****************************************************}
\typeout{****************************************************}
\typeout{}}

\end{article} 

\end{document}